\documentstyle[12pt]{article}

\begin{document}
\thispagestyle{empty}
\centerline{{\Large {\bf Reconnection rate for the steady-state
Petschek model}}}
\begin{center}
NIKOLAI V. ERKAEV \\[0pt]
{\sl Institute of Computational Modelling, Russian Academy of Sciences,
Krasnoyarsk, 66036, Russia }\\[0pt]
VLADIMIR S. SEMENOV \\[0pt] {\sl
Institute of Physics, State University of St. Petersburg, St. Petergof
198904, Russia }\\[0pt]
FERDINAND JAMITZKY\\[0pt]
{\sl Max-Planck-Institut f\"ur Extraterrestrische Physik,
P.O. Box 1603, 85740, Garching, Germany}\\[0pt]
\end{center}

\vspace{2cm}
\begin{abstract}
Reconnection rate is found for the canonical simplest
case of steady-state two-dimensional symmetric
reconnection in an incompressible plasma by matching of outer
Petschek solution and internal diffusion region solution.
The reconnection rate obtained naturally incorporates both
Sweet--Parker and Petschek regimes,
the latter seems to be possible only for the case with strongly
localized resistivity.
\end{abstract}

\newpage

{\large {\bf I. INTRODUCTION}}

Magnetic reconnection is an energy conversion process which occurs in
astrophysical, solar, space and laboratory plasmas (e.g., Hones$^1$;
Priest$^2$).
 First attempts to explain the fast energy release in solar flares
based on pure resistive magnetic field dissipation (Sweet$^3$; Parker$^4$)
showed that the energy conversion rate is estimated as
$1/\sqrt{Re_m}$, where
\begin{equation}
Re_m=\frac{V_A L}{\eta} \label{1}
\end{equation}
is the global Reynolds
number, $L$ is the half-length of reconnection layer, $V_A$ is
Alfv\'enic velocity, and $\eta$ is resistivity. For typical
conditions in the solar corona the Sweet-Parker rate turns out to be orders of
magnitudes too small when compared to experimental data.

In 1964 Petschek$^5$ pointed out that in a highly--conducting plasma
dissipation
needs only to be present  within a small region known as the diffusion
region, and energy conversion occurs primarily across non--linear waves,
or shocks. This gives another estimation of the maximum reconnection rate
$~1/\mbox{ln}Re_m$ which is much more favorable for energy conversion.

Unfortunately, up to the present it is still unclear which conditions
make Petschek-type reconnection to be possible and which are
responsible for the Sweet-Parker regime. The fact is that numerical
simulations (Biskamp,1986, Scholer, 1989) were not able to reproduce
solution of Petschek type but rather were in favor of Sweet-Parker
solution unless
the resistivity was localized in a small region (e.g., Scholer 1989, Yan,
Lee and Priest, 1992, Ugai, 1999).  The laboratory experiments also
seem to observe Sweet-Parker regime of reconnection (Uzdensky et al.,
1996, Ji et al.,1999).

From the mathematical point of view the problem of reconnection rate
is connected with the matching of a solution for the diffusion region
where dissipation is important, and solution for the convective zone
where ideal MHD equations can be used. But up to now this question is
still not resolved even for the canonical simplest
case of steady-state two-dimensional symmetric
reconnection in an incompressible plasma.

It is the aim of this paper to present a matching procedure for the
canonical reconnection problem. The reconnection rate obtained
from the matching turns out to incorporate
naturally both Petschek and Sweet-Parker regimes as limiting
cases.

\vspace{1cm}

{\large {\bf Petschek solution}}

We consider the simplest theoretical system consisting of a
two--dimensional current sheet which separates two
uniform and identical plasmas with oppositely oriented magnetic fields
$\pm {\bf B_0}$. Petschek (1964) pointed out that the diffusion region
can be considerably smaller than the whole size of the reconnection layer
and that the outer region contain two pairs of standing slow shocks.
These shocks deflect and accelerate the incoming plasma from the inflow
region
into two exit jets wedged between the shocks (see Figure 1). This jet area
between the shocks with accelerated plasma is traditionally called outflow
region.

In the dimensionless form the Petschek solution can be presented as follows
(Petschek, 1964, for details see Vasyliunas, 1975):

Inflow region:
\begin{equation}
v_x=0;\ v_y=-\varepsilon,  \label{2}
\end{equation}
\begin{equation}
B_x=1-\frac{4\varepsilon }\pi \ln \frac 1{\sqrt{x^2+y^2}};\quad B_y=\frac{
4\varepsilon }\pi \arctan \frac xy. \label{3}
\end{equation}
\noindent
Outflow region:
\begin{equation}
v_x=1;\ v_y=0;\ B_x=0;\ B_y=\varepsilon. \label{4}
\end{equation}
Equation of shock in the first quadrant is the following:
\begin{equation}
y=\varepsilon x. \label{5}
\end{equation}

Here $x,y$ are directed along the current sheet and in the perpendicular
direction, respectively. We normalized the magnetic field to $B_0$, length to
$L$, plasma velocity to Alfv\'enic velocity $V_A$, and electric field $E$
to Alfv\'enic electric field $E_A=V_A B_0$.

The reconnection rate
\begin{equation}
\varepsilon=E/E_A <<1 \label{6}
\end{equation}
is supposed to be a small parameter of the problem.

Expressions (\ref{2}-\ref{5}) are the asymptotic solution
with respect to $\varepsilon$ of the MHD system of
equations
\begin{eqnarray}
&(\bf {v \cdot \nabla)\bf v} = -{\bf
\nabla} P + ({\bf B \cdot \nabla)\bf B }, \label{7}
\\
&{\bf E} + ({\bf v}\times {\bf B}) = \frac{1}{Re_m}\mbox{curl}\bf {B},
\label{8}
\\
&\mbox{div}{\bf B}=0, \ \mbox{div}{\bf v}=0, \label{9}
\end{eqnarray}
and
the Rankine-Hugoniot
shock relations in the limit $Re_m\rightarrow
\infty$. Petschek did not obtain a solution in the diffusion region,
instead he estimated maximum
reconnection rate as $1/\mbox{ln}Re_m$
of using some simple physical suggestion . Generally speaking,
this implies that the Petschek model gives any reconnection rate from
Sweet-Parker value $1/\sqrt{Re_m}$ up to  $1/\mbox{ln}Re_m$, and
it is still unclear whether Petschek reconnection faster
than Sweet-Parker reconnection is possible.
The problem can be solved by matching of a solution for the diffusion
region and Petschek solution (\ref{2}-\ref{5}).

\vspace{1cm}

{\large {\bf Diffusion region scaling}}

We renormalize the MHD equations to the new scales
$B_0',\  V_A',\ E_A'=B_0' V_A'$,
where all quantities are supposed to be taken at
the diffusion region upper boundary, and at the half length of the
diffusion region $l_d$.  We have to use the dissipative MHD equations
(\ref{7}--\ref{9}) for the diffusion region with Reynolds number
\begin{equation}
 Re_m' = \frac {V_A' l_d} {\eta}, \label{10}
\end{equation}
and electric field $E=\varepsilon'$.

The scaling for the diffusion region is similar to that for
the Prandtl viscous layer (see Landau and Lifshitz, 1985):
\begin{eqnarray}
&x',\ B'_x,\ v'_x,\ P' \sim O(1), \nonumber
\\
&y',\ B'_y,\ v'_y, \ \varepsilon' \sim 1/\sqrt{Re_m'}. \label{11}
\end{eqnarray}
Consequently, the new boundary layer variables are the following:
\begin{eqnarray}
&\tilde x=x',\ \tilde B_x=B_x',\ \tilde v_x=v'_x,\ \tilde P=P', \nonumber
\\
&\tilde y=y'\sqrt{Re_m'},\ \tilde B_y=B_y'\sqrt{Re_m'},
\ \tilde v_y=v_y'\sqrt{Re_m'}, \
\tilde\varepsilon=\varepsilon'\sqrt{Re_m'}.\label{12}
\end{eqnarray}
The diffusion region Reynolds number is supposed to be $Re_m'>>1$,
and therefore in the zero-order with respect to the parameter
$1/ \sqrt{Re_m'}$ the boundary layer equations turn out to be:
\begin{eqnarray}
&\tilde v{_x} \frac{\partial \tilde v{_x}}{\partial \tilde x}+
 \tilde v{_y} \frac{\partial \tilde v{_x}}{\partial \tilde y}-
 \tilde B{_x} \frac{\partial \tilde B{_x}}{\partial \tilde x}-
 \tilde B{_y} \frac{\partial \tilde B{_x}}{\partial \tilde y}=
-\frac{\partial \tilde P(\tilde x)}{\partial \tilde x}, \label{13}
\\
&\mbox{div}{\bf \tilde B}=0,
\ \mbox{div}{\bf \tilde v}=0, \label{13a}
\\
&\tilde P=\tilde P(\tilde x), \label{14}
\\
& \tilde v{_y} \tilde B{_x}-
\tilde v{_x} \tilde B{_y}-\tilde \varepsilon=
\tilde \eta(\tilde x,\tilde y )
\frac{\partial \tilde B{_x}}{\partial \tilde y},  \label{15}
\end{eqnarray}
where $\tilde \eta(\tilde x,\tilde y)$ is the normalized resistivity of
the plasma with maximum value 1.

Unfortunately, the appropriate exact solutions of the boundary layer
equations (\ref{13}--\ref{15}) are unknown, therefore we have to solve the
problem numerically. The main difficulty is that the internal reconnection
rate $\tilde \varepsilon$ is unknown in advance and has to be determined
for given resistivity $\tilde \eta(\tilde x,\tilde y)$, given total
pressure $\tilde P(\tilde x)$, and $\tilde B_x(\tilde x)$
given at the upper boundary of the
diffusion region. In addition, the solution must have
Petschek-type asymptotic behaviour (\ref{2}--\ref{5}) outside of the diffusion
region.

Although we are looking for a steady-state solution,
from the simulation point of view it is advantageous to use relaxation
method and solve numerically the following unstationary system of
boundary layer MHD equations:
\begin{eqnarray}
&\frac{\partial \tilde v}{\partial t}+
\tilde v{_x} \frac{\partial \tilde v{_x}}{\partial \tilde x}+
 \tilde v{_y} \frac{\partial \tilde v{_x}}{\partial \tilde y}-
 \tilde B{_x} \frac{\partial \tilde B{_x}}{\partial \tilde x}-
 \tilde B{_y} \frac{\partial \tilde B{_x}}{\partial \tilde y}=
-\frac{\partial \tilde P(\tilde x)}{\partial \tilde x}, \label{16}
\\
& \frac{\partial {\bf \tilde B}}{\partial t} =
\mbox{curl}({{\bf \tilde v\times \tilde B}})-
\mbox{curl}\left( \eta(\tilde x,\tilde y)\mbox{
curl} {\bf \tilde B}\right),
\label{17}
\\
&\mbox{div}{\bf \tilde B}=0,
\ \mbox{div}{\bf \tilde v}=0. \label{18}
\end{eqnarray}
Starting with an initial MHD configuration under fixed boundary conditions
we look for convergence of the time-dependent solutions to a steady state.

As initial configuration we choose a X-type flow and magnetic field:
$\tilde v_x=\tilde x,\ \ \tilde v_y=-\tilde y,\ \
 \tilde B_x=\tilde y,\ \ \tilde B_y=-\tilde x$. The
distribution of the resistivity is traditional (see (Ugai,1999, Scholer
1985)):
\begin{equation}
\eta(\tilde x,\tilde y)=d e^{(-s_x \tilde x^2-s_y \tilde y^2)}+f,
\label{19}
\end{equation}
with $d+f=1$ where coefficient $d$ describes inhomogeneous resistivity,
and $f$ is responsible for the background resistivity.

The problem under consideration consists essentially of two coupled
physical processes: diffusion and wave propagation. To model these
processes, two-step with respect to time numerical scheme has been used.
At first, convectional terms were calculated using the Godunov characteristic
method, and then the elliptical part was treated implicitly.

Calculations were carried out on a rectangular uniform grid
$100\times145$ in the first quadrant
with the following boundary conditions:

\noindent
Lower boundary: symmetry conditions
$\partial \tilde v_x/\partial y=0,\ \ \tilde v_y=0,\ \  B_x=0$;
induction equation (\ref{17}) has been used to compute
the $B_y$ component at the $x$--axis.

\noindent
Left boundary: symmetry conditions
$\tilde v_x=0,\ \
\partial \tilde v_y/\partial x=0,\ \
\partial \tilde B_x/\partial x=0,\ \
\tilde B_y=0$.

\noindent
Right boundary: free conditions
$\partial \tilde v_x/\partial x=0,\ \
\partial \tilde v_y/\partial x=0$.

\noindent
Upper (inflow) boundary:
$\tilde v_x=0, \ \ \tilde B_x=1$.

Note, that this implies that we do not prescribe the incoming velocity,
and hence the reconnection rate: the system itself has to determine how
fast it wants to reconnect.

The total pressure can be fixed to 1 in the zero-order approximation:
$\tilde P=1$.

Let us discuss the result of our simulations. For the case of localized
resistivity where we chose $d=0.95,\ \ f=0.05,\ \ s_x=s_y=1$ in the
equation (\ref{19}), the system
reaches Petschek steady state (see Figure 2) with clear asymptotic
behaviour, pronounced slow shock, and the
reconnection rate turns out to be $\tilde\varepsilon \sim 0.7$.

From the other hand, for the case of homogeneous resistivity
 $d=0,\ \ f=1$, the system reaches Sweet-Parker state (see Figure 3)
with much less reconnection rate $\tilde\varepsilon \sim .25$ even
if the Petschek solution has been used as initial configuration
(see also (Ugai, 1999,
Scholer,1989)). This seems to imply that Petschek-type reconnection
is possible only if the resistivity of the plasma is localized
in a small region, and for constant resistivity the
Sweet-Parker regime is realized.

The size of the diffusion region $l_d$ can be defined as the
size of the region where the convective electric field
$E=v\times B$ (which is zero at the
origin) reaches the asymptotic value
$\tilde\varepsilon$ (or, some level, say $0.95\tilde\varepsilon$).
For the case of localized resistivity $l_d$ practically coincides with the
scale of the inhomogeneity of the conductivity.  In principal, there might be
a possibility to produce Petschek-type reconnection with constant
resistivity using a highly inhomogeneous behaviour of the MHD parameters at the
upper boundary (narrow stream, for example, see Chen et al.,1999), and
then $l_d$ has the meaning of the scale of this shearing flow or other
boundary factor which causes the reconnection.

\vspace{1cm}

{\large {\bf  Matching procedure}}

We have only a numerical solution for the diffusion region, and this makes
it difficult for the matching procedure because the latter needs an analytical
presentation of the solutions to be matched. The only way out left is to
continue the diffusion region solution to the inflow region using dates known
from the simulation distribution of the $B_y$ component along the upper
boundary of the diffusion region. Then try to match the solutions in the
current free
inflow region at the distance $r\sim l_d$ (see Figure 1).

As can be seen from equation (\ref{3}) the
$B_x$ component of the Petschek solution diverges
at the origin $B_x\rightarrow
-\infty $ when ${r=\sqrt{x^2+y^2}}\rightarrow 0$. This singularity is a
consequence of the fact that dissipation actually has not been taken into
account for the solution (\ref{2}- \ref{5}) which is nevertheless still valid
untill the distances of the order of the size of diffusion region is $l_d$.

In order to be adjusted to the Petschek solution, the
$B_y'$ component must have the
following limit for $x/l_d\rightarrow\infty$
at the upper boundary of the diffusion region :
\begin{equation}
B_y'(x/l_d)\rightarrow 2\varepsilon. \label{19a}
\end{equation}
We can obtain the asymptotic behaviour of $B_x'$ for $r>l_d$ region using a
Poisson-like integral presentation:
\begin{eqnarray}
&\displaystyle B_x'(x',y')=
B_0'+\frac 1\pi \int_{-\infty}^{+\infty}
\frac{\partial B_y'^{(1)}(\tilde{x},0)}{\partial x}
\ln{\frac{{\sqrt{(x-\tilde{x})^2+y^2}}}{l_d}}d\tilde{x}= \nonumber
\\
&\displaystyle B_0'+\frac 1\pi \int_{-\infty}^{+\infty}
\frac{\partial B_y'^{(1)}(\xi,0)}{\partial \xi} \left\{
\ln{\frac{{\sqrt{x^2+y^2}}}{l_d}}+
\frac{\xi^2-2x\xi}{x^2+y^2}\right\}d\xi= \nonumber
\\
&\displaystyle B_0'+
\frac{4\varepsilon }{\pi} \ln \frac{r}{l_d}+O(1/r),  \label{20}
\end{eqnarray}
where $\xi=x/l_d$.
This gives an outer expansion for the inner solution.
On the other hand a convective solution (\ref{3})
can be rewritten in the following form in order
to determine the inner expansion of the outer solution:
\begin{equation}
B_x=1-\frac{4\varepsilon }\pi \ln \frac Lr=
1-\frac{4\varepsilon}\pi \ln \frac L{l_d}-
\frac{4\varepsilon}\pi \ln \frac {l_d}r. \label{21}
\end{equation}

Equating these two asymptotic expansions we obtain the matching relation:
\begin{eqnarray}
B_0'=1-\frac{4\varepsilon}\pi \ln \frac L{l_d}, \label{22}
\end{eqnarray}
Now everything is ready to determine the reconnection rate. The electric field must
be constant in the whole inflow region, hence
\begin{eqnarray}
&v'B_0'=vB_0, \label{23}
\\
&\varepsilon' B_0'^2 = \varepsilon B_0^2 , \label{24}
\end{eqnarray}
where the definition of the reconnection rates $\varepsilon'=v'/B_0',\ \
\varepsilon=v/B_0$ has been used.
Bearing in mind that $\varepsilon '=\tilde\varepsilon/\sqrt Re_m'$
(see scaling (\ref{12}) we obtain:
\begin{equation}
\tilde\varepsilon{B_0'}^{3/2}=
\varepsilon B_0^{3/2} \sqrt{\frac{l_dB_0}{\eta}}. \label{25}
\end{equation}

Substituting $B_0'$ from the equation (\ref{22}) we determine finally the following
equation for the reconnection rate $\varepsilon$ :
\begin{equation}
\tilde\varepsilon(1-\frac{4\varepsilon}\pi \ln \frac L{l_d})^{3/2}=
\varepsilon \sqrt{Re_m\frac{l_d}{L}}, \label{26}
\end{equation}
where $Re_m$ is the global Reynolds number (\ref{1}), and the internal
reconnection rate $\tilde\varepsilon$ has to be found from the
simulation of the diffusion region problem.

For small $\varepsilon$ there is an analytical expression:
\begin{equation}
\varepsilon=\frac{\tilde\varepsilon}{\sqrt{Re_m\frac{l_d}{L}}+
\frac 6\pi \tilde\varepsilon\ln \frac L{l_d}}. \label{27}
\end{equation}
Here $ \tilde\varepsilon$ is an internal reconnection rate  determined
from the numerical solution: $\tilde\varepsilon \sim 0.7$.
\vspace{1cm}

{\large {\bf Discussion and conclusion}}

Equations (\ref{26},\ref{27}) give the unique reconnection rate for known
parameters of the current sheet $L,\ \ B_0,\ \ V_A,\ \ \eta,\ \ l_d$.
For sufficiently long diffusion region such that
$\sqrt{Re_m\frac{l_d}{L}}>> \frac 6\pi \tilde\varepsilon\ln \frac L{l_d}$,
the equation (\ref{27}) corresponds to Sweet-Parker regime $\varepsilon
\sim  \tilde\varepsilon/\sqrt{Re_m \frac{l_d}{L}}$.
 For the opposite case of resistivity
constrained in a small region $\varepsilon \sim \frac \pi 6 /\ln \frac L{l_d}$
we have Petschek reconnection. Hence, reconnection rate
 (\ref{26},\ref{27}) naturally incorporates both regimes obtained in
simulations (Scholer, Ugai, Biskump).

We were not able to reproduce Petschek regime using variation of MHD
parameters at the upper
boundary with homogeneous resistivity, a probably solution (Chen, 1999) of this problem
either is essentially time-dependent or corresponds to the case of strong
reconnection.
 According to our simulations, for Petschek state to
exist a strongly localized resistivity is needed, and for the spatially
homogeneous resistivity $l_d=L$ Sweet--Parker regime seems to be always
the case. This result resolves old question about conditions which are
necessary for Petschek-type reconnection to appear.

It is interesting that for the deriving of equations
(\ref{26},\ref{27}) the only value which has been actually used is
the internal reconnection rate $\tilde\varepsilon$
obtained from the numerical solution, but the distribution
of the $B_y$ component along the upper boundary of the diffusion region
does not contribute at all (besides asymptotic behaviour (\ref{20}))
in the zero--order approximation considered above. Of course,
from the mathematical point of view it is
important that diffusion region solution exists and has Petschek--like
asymptotic behaviour (\ref{2}--\ref{4}).

The strongly localized resistivity is often the relevant case in space
plasma applications, but for the laboratory experiments where the size
of a device is relatively small the Sweet--Parker regime is expected.
\vspace{1cm}

{\large {\bf VIII. ACKNOWLEDGEMENTS}}

We thank M.Scholer, M. F. Heyn and H. K. Biernat for useful discussions
and help.
VSS was supported by the Russian Foundation for Basic Research --
Deutsche Forschungsgemeinschaft, grant \mbox{98--05--04073.}
NVE was supported in part by grant No 98-05-65290 from Russian
Foundation of Basic Research and by Russian grant No 97-0-13.0-71
from Russian Ministry of Education.

\vspace{1cm}

{\large {\bf IX. REFERENCES}}

\everypar={\hangafter=1\hangindent=10mm}

E. W. Hones,  Jr.,{ \it Magnetic Reconnection in Space and
Laboratory Plasmas} (Geophysical Monograph 30, AGU, Washington, 1984).

E. R. Priest,  Rep.
Progr. Phys., {\bf 48}, 955 (1985).

P. A. Sweet, in {\it
Electromagnetic
Phenomena in Cosmic Physics}, edited by B. Lehnert (Cambridge
University Press, London, 1958),  p. 123.

E. N. Parker,
Astrophys. J. Suppl. Ser., {\bf 8}, 177
(1963).

H. E. Petschek, in {\it AAS--NASA Symposium of the Physics of Solar
Flares, NASA--SP 50}, edited by W. N. Hess (National Aeronautics and Space
Administration, Washington, DC, 1964), p.  425.

V. M. Vasyliunas,
Rev. Geophys. Space Phys., {\bf 13}, 303 (1975).

D. Biskamp, Magnetic reconnection via current sheets, Phys. Fluids,
29, 1520, 1986.

Scholer, M., Undriven reconnection in an isolated current sheet,
J.Geophys. Res.,94, 8805, 1989.

Yan, M., L.C.Lee and E.R.Priest, Fast magnetic reconnection with small
shock angles, J.geophys.Res., 97, 8277, 1992.

Ugai,M., Computer studies on the spontaneous fast reconnection model as a
nonlinear instability, Phys. Plasmas,6, 1522, 1999.

Uzdensky D.A., R.M.Kulsrud, and M. Yamada, Phys.Plasmas, 3, 1220, 1996.

 L. D. Landau  and E. M. Lifschitz, {\it Klassische Feldtheorie}
(Akademie--Verlag, Berlin, 1984).

Chen, T., Z. X. Liu, and X. X. Zhang, Transient reconnection caused by the
impact and switch-off of a transverse shear flow, Phys. Plasmas,
6, 2393, 1999.

Ji, H., M. Yamada, S. Hsu, R. Kulsrud, T. Carter, and S. Zaharia,
Magnetic reconnection with Sweet--Parker characteristics in
two-dimensional laboratory plasmas, Phys. Plasmas, 6, 1743, 1999.


\newpage

{\large {\bf  Figure Captions }}

Figure 1:
{Scheme of matching of the outer Petschek solution and diffusion region
solution.}

Figure 2:
{Configuration of magnetic field lines (solid line) and stream lines
(dashed line) for the numerical simulation of the diffusion region.}

Figure 3:
{Three-dimensional plot of current density shows Petschek shock}

\vfill

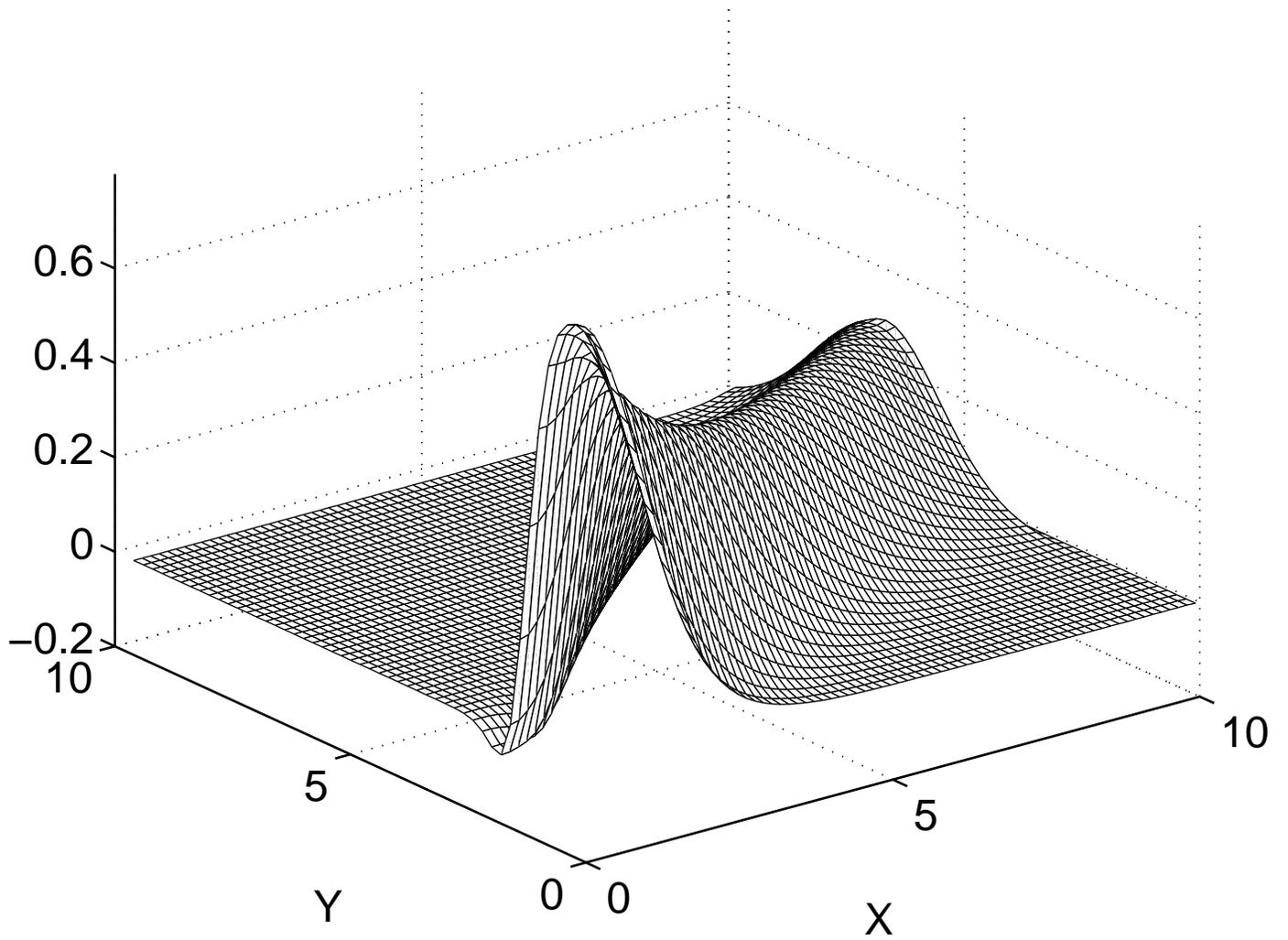
\end{document}